\documentclass[prd, twocolumn, nofootinbib, floatfix]{revtex4}

\usepackage{amsmath}
\usepackage{amsbsy}

\input epsf
\usepackage{amsmath,amssymb,subfigure}
\usepackage{graphicx}
\usepackage{epsfig}
\usepackage{color}
\usepackage{ulem}
\usepackage{epstopdf}

\newcommand{\ls}{\mathrel{\raise0.27ex\hbox{$<$}\kern-0.70em \lower0.71ex\hbox{{
$\scriptstyle \sim$}}}}

\begin{document}

\title{WMAP 7 Constraints on Oscillations in the Primordial Power Spectrum}

\author{P.~Daniel Meerburg$^{(1,2)}$, Ralph Wijers$^{(1,2)}$,  and Jan Pieter van der Schaar$^{(1,3)}$}
\affiliation{$^1$ Gravitation and AstroParticle Physics Amsterdam,  University of Amsterdam, Science Park 904, 1098XH Amsterdam, The Netherlands\\
$^2$ Astronomical Institute ``Anton Pannekoek", University of Amsterdam, Science Park 904, 1098XH Amsterdam, The Netherlands\\
$^3$ Korteweg-de Vries Institute for Mathematics, University of Amsterdam, Science Park 904, 1098XH Amsterdam, The Netherlands}

\date{\today}

\begin{abstract} 
We use the WMAP 7 data to place constraints on oscillations supplementing an almost scale-invariant primordial power spectrum. Such oscillations are predicted by a variety of models, some of which amount to assuming there is some non-trivial choice of the vacuum state at the onset of inflation. In this paper we will explore data-driven constraints on two distinct models of initial state modifications. In both models the frequency, phase and amplitude are degrees of freedom of the theory for which the theoretical bounds are rather weak: both the amplitude and frequency have allowed values ranging over several orders of magnitude. This requires many computationally expensive evaluations of the model CMB spectra and their goodness-of-fit, even in a Markov Chain Monte Carlo (MCMC), normally the most efficient fitting method for such a problem. To search more efficiently we first run a densely spaced grid, with only 3 varying parameters; the frequency, the amplitude and the baryon density. We obtain the optimal frequency and run an MCMC at the best fit frequency, randomly varying all other relevant parameters. To reduce the computational time of each power spectrum computation, we adjust both comoving momentum integration and spline interpolation (in $l$) as a function of frequency and amplitude of the primordial power spectrum.  Applying this to the WMAP 7 data allows us to improve existing constraints on the presence of oscillations. We confirm earlier findings that certain frequencies can improve the fitting over a model without oscillations. For those frequencies we compute the posterior probability, allowing us to put some constraints on the primordial parameter space of both models. 
 \end{abstract}



\maketitle
\section{Introduction}

The observed statistical distribution of temperature fluctuations in the cosmic microwave background (CMB) is believed to be largely determined by the physics in the very early Universe. These CMB fluctuations were sourced by quantum fluctuations during an epoch of accelerated expansion early on in the history of the Universe, known as inflation. They then induce curvature perturbations in the geometry of spacetime, which are preserved after horizon crossing during inflation. Once they re-enter the horizon at some later time, they couple to radiation and matter, becoming responsible for the observed statistical distribution of the CMB temperature fluctuations and the large scale structure (LSS) of matter in the Universe, respectively. Within the 6-parameter $\Lambda$CDM these initial conditions are described by only 2 parameters: the amplitude of the primordial power spectrum of scalar perturbations, and the tilt $n_s$ describing the scale-dependence of the power spectrum to first order. By investigating the CMB power spectrum, one is therefore able to probe high energy physics in the early Universe. Given the large number of models describing the physics of the early Universe (see e.g., \cite{Chen:2010xka} for a recent overview), these two parameters (within 6-parameter $\Lambda$CDM) are not sufficient to be able to discriminate between various proposed models. Additional degrees of freedom, derived from the statistical analysis of the late time distribution of temperature (CMB) or density (LSS) fluctuations, could potentially be used to break the degeneracy between various models. Possible extensions to the 6-parameter model include tensor degrees of freedom (gravitational waves), higher order corrections to the scalar {\it and} tensor power spectra, and deviations from Gaussianity measured through high order correlation functions. The ultimate goal, of course, is to formulate a theoretical prediction of what these degrees of freedom should be, and constrain these from the observational data, just as we constrain the 2 parameters describing the tilt and the amplitude of the primordial power spectrum of scalar perturbations. The additional constrained parameters test the uniqueness of a proposed model and thus contribute to the understanding of the physics governing the early, inflationary Universe. 

In this paper we will consider modifications to the primordial power spectrum. In particular, we will search for evidence of oscillations in the almost scale-invariant spectrum. The motivation to search for these modifications is provided by a rapidly increasing number of theoretical models expecting such features in the primordial correlation statistics: e.g., in the power spectrum \citep{Martin:2000xs, Easther2001a, Easther2003a, Danielsson2002,EKP2,SSS2004,Schalm:2004xg,GSSS2005,Achucarro:2010da, Cai:2010uw,Flauger:2009ab,Chen:2010bka,Jackson:2010cw,Chen:2011zf,Jackson:2011qg} and bispectrum \citep{Chen2007a,Chen2008, Meerburg2009,Flauger:2010ja, Meerburg2010a,Chen:2010bka}. In this paper we will only consider oscillations in the power spectrum. \cite{Fergusson:2010dm} have attempted to constrain features in the bispectrum and \cite{Meerburg2010b} have proposed a method to effectively search for features in the CMB bispectrum; recently LSS data has been proposed to search for these type of bispectra \citep{CyrRacine:2011rx}. The frequency, amplitude, and possibly the phase and first order scale dependence of these features are determined by the detailed physics of inflation. Detecting or constraining these parameters would help us determine the precise physics of inflation. 

\begin{figure}
\begin{centering}
\includegraphics[scale=.45]{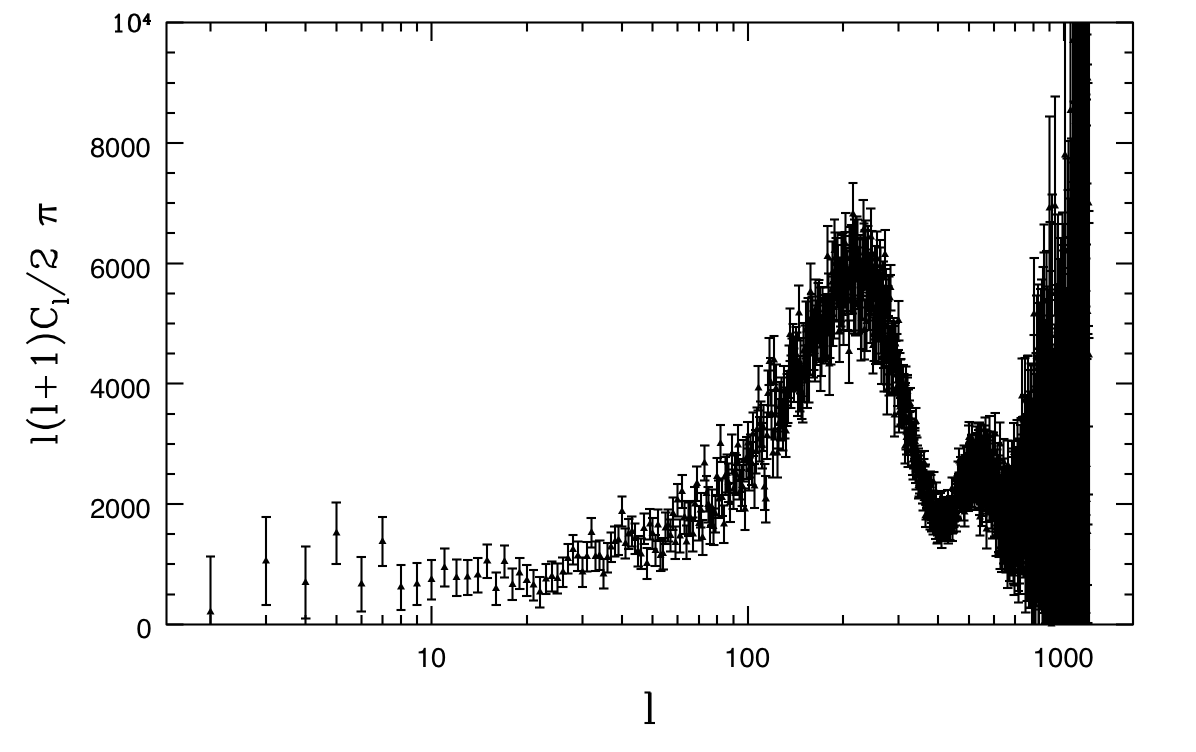}
\par\end{centering}
\caption{WMAP 7 data with added errors from measurement and cosmic variance. The error bars are derived from the diagonal terms in the Fisher matrix. The multipole moments are slightly coupled (inducing off-diagonal elements), so a correct treatment of errors requires use of the entire Fisher matrix, which is done when calling the likelihood WMAP code.}
\label{fig:Clerror}
\end{figure}

This is not the first time features in the power spectrum will be explored. Notably \cite{EKP1,EKP2}, \cite{Martin:2003sg,Martin:2004yi}, \cite{Covi:2006ci}, \cite{Hamann:2007pa}, \cite{Pahud2009}, and \cite{Flauger:2009ab} have all investigated possible oscillations in the primordial power spectrum. Although there has been no clear detection, the data certainly allow small oscillations, as can be seen from fig.~\ref{fig:Clerror}. Previous analyses were done on WMAP 1-, 3-, and 5-year data (as well as SDSS data). In this paper we will aim at extending and improving the analysis using the latest WMAP release, WMAP 7 \citep{Komatsu:2010fb}. One of the key hurdles in searching for oscillations in the data is the frequency of the oscillations. The frequency of the oscillations in the primordial power spectrum is a free parameter which spans several orders of magnitude for many of the proposed models. When probing the joint likelihood of our cosmological model, the large range in frequencies results in a number of issues \citep{Groeneboom:2007rf}. 

Foremost, it requires a high resolution in `sample space', i.e., looking for the best fit parameter to the data requires a large number of computations of the CMB power spectrum from the primordial power spectrum (using, e.g., CAMB \citep{Lewis:1999bs} or CMBfast \citep{Zaldarriaga}). This necessitates an efficient computing scheme. Unfortunately, the computation of the late time power spectrum from the primordial one involves a convolution of the transfer function $\Delta_l(k)$ with the primordial power spectrum $P(k)$. The appearance of rapid oscillations in both the transfer functions as well as the primordial spectrum requires smaller and more frequent steps in the integration in comoving momentum space $k$, increasing the required computational time for each run significantly. 

Secondly, the late time spectrum must to be computed for each angular scale $l$. Usually, given the smoothness of the primordial spectrum, it suffices to do a spline interpolation on a number of $C_l$. This reduces the computational time, since the computation of the transfer functions\footnote{These can generally not be precomputed, as they depend on all relevant parameters of the theory, which are varied when searching for the best fit.} is the most time-consuming part. With the addition of oscillations on top of the smooth primordial spectrum, the number of $C_l$ necessary to obtain an accurate fit of the interpolated $C_l$ will depend on the frequency of the primordial oscillations. As this frequency increases, at some point {\it all} $l$ will need to be considered in order to resolve the superimposed oscillations. Computing all $l$ requires us to compute all transfer functions, which again increases the computational time. Lastly, there will often be a number of frequencies that tend to improve the fit within the large range we explore, rather than just a single one. For example, if a frequency $\omega$ is a good fit to the data, there is a fair chance $2 \omega$ will be a good fit as well. This is a major issue, as a multidimensional parameter space is most effectively searched through a Markov Chain  Monte Carlo (MCMC), which is a random process. Therefore, when the frequency is not fixed, an MCMC approach to the best fit is not efficient as the likelihood is spiked and the random nature of the MCMC chain will lead to frequent tunneling of one local maximum to another within the multidimensional joint likelihood.  \cite{Flauger:2009ab} showed a way to circumvent this issue by first taking a large number of samples fixed on a grid. Obviously a grid does not allow us to vary all parameters within our cosmological model, as the number of samples grows quickly with the dimensionality of parameter space. Instead, the priority lies in varying the initial conditions, i.e., the oscillatory component of the primordial power spectrum. Once the best fit has been determined (within the prior frequency range, and the grid resolution), one can run MCMCs with a fixed best fit frequency determined through the grid. The joint likelihood is expected to no longer contain local maxima and an MCMC should converge quickly.  This eventually allows us to put constraints on the amplitude and perhaps the first-order scale dependence of the amplitude of the oscillatory feature.

Although there are a large number of different features and oscillations predicted by a variety of models, in this paper we will focus on only two, distinct, theoretically motivated modified primordial power spectra. We will introduce these modified power spectra in section \ref{sec:models}. In section \ref{sec:codes1} we explain how to optimize the search by making the numerical computation of the power spectrum frequency-dependent. To find a best fit frequency before we apply the MCMC, we use grid sampling. We report our findings in section \ref{sec:gridsampling}. Once we have established a best fit value of the frequency we run an MCMC with that best fit value for two models. The results are discussed in section \ref{sec:MCMC} and we compare these to theoretically derived constraints on primordial parameter space. We conclude in section \ref{sec:conclusions}.

\section{Two models of initial state
modifications}\label{sec:models}

Although the standard BD vacuum state during inflation is an excellent
fit to the currently available CMB data, theoretical considerations have
questioned its validity and uniqueness over the last decade (see
\cite{Martin:2000xs, Easther2001a, Easther2003a, Danielsson2002,EKP2,SSS2004,Schalm:2004xg,GSSS2005} and references therein). The main
reason for casting doubt on the BD assumption is the fact that the
temperature fluctuations in the CMB ultimately find their origin in
quantum fluctuations in a vacuum state right at the onset of inflation.
Predicting the spectrum of CMB temperature fluctuations requires
determining this initial vacuum state for quantum modes at
extraordinarily high momentum scales, far beyond any high-energy cut-off
scale, where a perturbative quantum field theory description is no
longer expected to be accurate. From a theoretical point of view it is
not understood why the time-dependent inflationary background would
transfer this unknown high-energy physics to a regular quantum field
theory description in the BD state at later times. In wait for a more
complete understanding this vacuum state ambiguity has motivated
phenomenological approaches, usually relying on low-energy effective
field theory expectations, that have suggested the presence of
high-energy corrections to the BD state. A typical prediction of these
models is the appearance of (small) oscillations on top of the standard
primordial power spectrum of inflationary perturbations.

Besides constraining these characteristic oscillations in general, an
additional goal of this work is to study to what extent the currently
available CMB power spectrum data can distinguish between two rather
general classes of models that have been proposed to describe initial
state modifications.  The first class of models is known as the Boundary
Effective Field Theory (BEFT) approach to determine initial-state
modifications \citep{SSS2004, GSSS2005}.  In this proposal one fixes an
initial {\it time} where one calculates corrections to the usual BD
initial condition using a low-energy effective boundary Lagrangian. The
result is an explicitly scale-dependent Bogolyubov parameter $\beta_{\bf
k}$ describing the modification with respect to the usual BD vacuum
state, giving rise to strongly scale dependent oscillatory corrections
to the primordial power spectrum.

The second class of models is known as the New Physics Hypersurface
(NPH) approach to initial-state modifications \citep{Danielsson2002,
Easther2002}. In the NPH scenario one traces every momentum mode back to
some large {\it physical scale} of new physics $M$ and imposes, on a
rather ad-hoc basis, the standard flat space vacuum state (corresponding
to positive frequency modes only), mode by mode, resulting in a $k$
independent Bogolyubov parameter $\beta_{\bf k}$. A small departure from
scale-invariance only arises after taking into account the slow-roll
evolution of the Hubble parameter, which affects the amplitude of the
oscillatory corrections described by $\beta_{\bf k} \propto H/M$. A
recent effort by Jackson and Schalm \cite{Jackson:2011qg} to compute the
low-energy vacuum state effects of a massive scalar field in an
inflationary background carries the important prospect to ground the NPH
proposal in a more solid effective field theory description. The authors
showed that integrating out an arbitrary massive scalar field can indeed
affect the vacuum, resulting in rather similar corrections to the
inflationary power spectrum as in the original NPH scenario. For our
purposes the power spectrum predicted in \cite{Jackson:2011qg} will be
used to set-up a general phenomenological parameterization of the almost
scale-invariant NPH class of models.

These two classes of models describing initial state modifications, NPH
and BEFT, serve as a nice benchmark to study how constraining and
distinguishing the most recent CMB power spectrum data is when
considering superimposed oscillations.
The data analysis is in fact best performed using a primordial power
spectrum parameterization of the oscillating corrections involving a set
of independent parameters. The general parameterization of the
primordial power spectrum for the two classes of models that we will
make use of are \citep{GSSS2005}
\begin{eqnarray}
P(k) &=& P_0(k) \left[1 + \beta k A_0 \sin \left( 2 A_1 k +\phi \right) \right]
\label{eq:PK_beft}
\end{eqnarray}
for the BEFT scenario, while for the new Jackson and Schalm scenario \citep{Jackson:2011qg} (from here on NPH) 
\begin{eqnarray}
P(k) &=& P_0^*\left[ a_0 +a_1 \ln k/k_* + \right.\nonumber\\
&& \left.(a_2 +a_3 \ln k/k_*)\sin \left( a_4 \ln k/k_* +\zeta\right) \right].
\label{eq:PK_nph}
\end{eqnarray}
In eq.~\eqref{eq:PK_beft} $\beta$ is a parameter determined by the
details of the physics in the ultraviolet, and is naturally expected to
be $\mathcal{O}(1)$, while $A_0$ and $A_1$ are (related) parameters
predicted by the BEFT method. $P_0(k)$ is the power spectrum from
canonical slow-roll inflation in a BD state. This contains a possible
tilt $n_s$. In eq.~\eqref{eq:PK_nph}, $P_0^*$ is the scale independent
power spectrum, and a possible tilt is not included. The reason for this
difference is that the oscillatory correction in the NPH scenario can
have a tilt $a_3$ that differs from the tilt $a_1 = n_s-1$. For the BEFT
we will be interested in possibly constraining $A_0$, $A_1$ and $\phi$,
while for the NPH model we will consider $a_2$, $a_3$, $a_4$ and
$\zeta$.  The parameters $a_0$ and $a_1$ will be considered zero-order
contributions to the power spectrum and are constrained through $n_s$
and $P_*$. As already briefly alluded to, we will constrain each
parameter independently and apply theoretical dependencies between those
parameters only afterwards. For example, $A_1$ in the BEFT model is
related to the amplitude $A_0$. The reason not to implement this
directly is because separating these parameters results in a much
`cleaner' sampling, because the frequency causes the likelihood to be
highly irregular. We will apply the theoretically predicted relations
between the amplitude and the frequency once the posterior distribution
has been determined.

For more specific details and theoretical background on the NPH and BEFT
models we refer to \cite{GSSS2005} and \cite{Jackson:2011qg,
Danielsson2002}, respectively.

\section{Code Optimization}\label{sec:codes1}


We have modified the CAMB code \citep{Lewis:1999bs} in order to efficiently search for oscillations. The modification is built upon work done in \cite{Martin:2003sg,Martin:2004yi}. The late time power spectrum basically is a convolution of the transfer function $\Delta_l(k)$ and the primordial power spectrum $P(k)$
\begin{eqnarray}
C_l &\propto& \int_0^{\infty} k^2 dk P(k) \Delta^2_l(k).
\end{eqnarray}
The transfer function contains all the physics that governs the evolution of the Universe and describes an interplay between radiation pressure and gravitational collapse. This causes the transfer function to be highly oscillatory. Consequently, to gain sufficient accuracy one has to numerically integrate over $k$-space with an adequate number of steps. The transfer functions are not only a function of comoving momentum, but also depend on the angular scale $l$. To increase the speed of the code, one usually does not compute the complete transfer matrix, but rather limited number of $l$, and spline interpolates between them. 

If the primordial power spectrum is smooth both the integral over comoving momentum and the spline interpolation in $l$ can be done with limited samples. Once we allow for (rapid) oscillations in the primordial power spectrum, the number of samples needs to be increased. 

Numerically, the most time consuming application is the number of transfer functions we will have to compute. Therefore the fewer transfer functions we compute, the faster we can determine the angular power spectrum $C_l$. Obviously, if the number of oscillations in
our primordial spectrum is large, we will need many transfer functions to resolve those using a spline interpolation. If the wavelength of the primordial signal is $\delta l\sim1$, no spline interpolation will resolve the primordial signal.

\subsection{Frequency-Dependent Power Spectrum Computation}\label{sec:codes2}

\begin{figure}
\begin{centering}
\includegraphics[scale=.45]{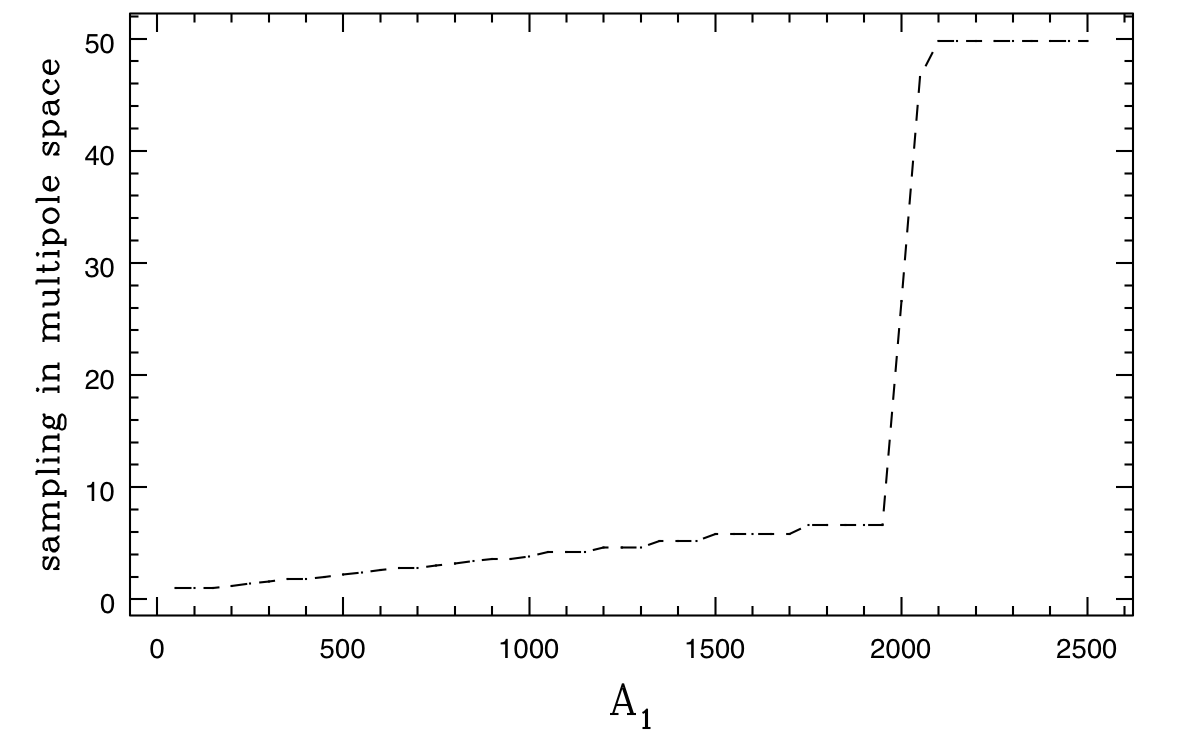}
\par\end{centering}
\caption{Sampling increase for splining in multipole space as a function of $A_1$ for the BEFT model. We fixed $A_0 = 10$, its largest possible value. As the frequency $A_1$ increases, it requires an increasing number of neighboring $l$ in the interpolation to maintain the same accuracy (which was set by WMAP best-fit standards). Around $A_1\sim 2000$ the sampling factor increases beyond 50, which implies that all $l$'s are required (i.e., splining is not sufficient and one needs to compute the full transfer matrix). }
\label{fig:sbBEFT}
\end{figure}

We have modified CAMB and COSMOmc in order to optimally compute the power spectrum given a primordial frequency. Fixing all parameters to the best fit WMAP 7 values \citep{Komatsu:2010fb}, except for the frequency in the primordial power spectrum, we investigated the convergence in the late time power spectrum when we increased the sampling in $k$- and $l$-space. As a null hypothesis, with `optimal' accuracy,  we took the primordial spectrum without oscillations. By plotting the frequency against the sample increase required to retain the same accuracy, we obtained an estimate of the optimal sampling for every frequency. As an example, in fig.~\ref{fig:sbBEFT} we show the increase in $l$-sampling required to obtain an accurate computation of the power spectrum as a function of the BEFT frequency $A_1$. Only for relatively low frequencies $A_1<2000$ and $a_4 <60$ can we use splines in $l$-space to determine the CMB power spectrum. At higher frequencies we need to compute the full transfer matrix in order to resolve oscillations\footnote{The sampling is from a set of predetermined $l$ (e.g., $l=\{1,2,3,....100,150,...1000,1100,...\}$). The number of points used in the interpolation is increased within this set of predetermined $l$ until this set is unable to achieve the same accuracy as one would obtain by computing all $l$. This set is based on efficient computation of a smooth primordial spectrum.}. For the sampling in $k$-space the optimal sampling was derived when computing all transfer functions. We find that over the range of frequencies we will probe in this paper we need to approximately double the resolution in $k$-space when doing the integration.

\section{Grid Sampling} \label{sec:gridsampling}
\begin{figure*}
\begin{centering}
\includegraphics[width=\textwidth]{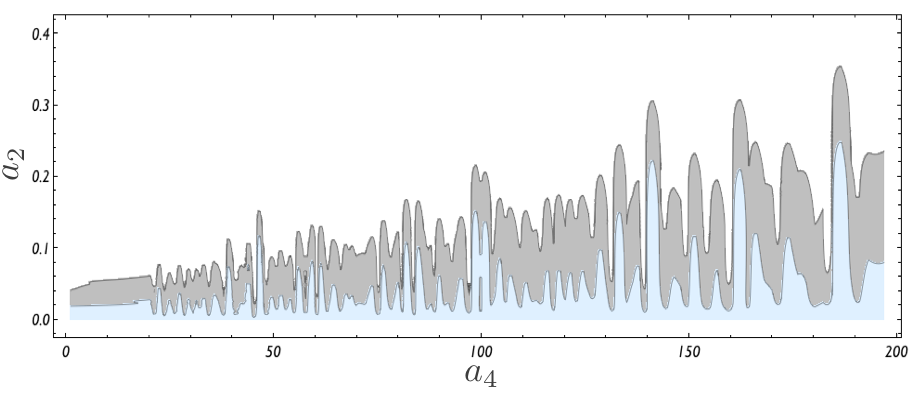}
\par\end{centering}
\caption{(NPH) The $68\%$ and $95\%$ confidence levels for amplitude $a_2$ versus the frequency $a_4$ for the low frequency grid $1\leq a_4 \leq 200$ marginalized over the baryon density. There are many local peaks in the marginalized likelihood. The two most likely grid points are ($a_4 = 46.5$, $a_2 = 0.056 $) and ($a_4 = 98.41$, $a_2 = 0.147 $). Towards higher frequencies, larger amplitudes are allowed by the data but do not necessarily represent significant improvements of the overall fit.}
\label{fig:gridcontournph2}
\end{figure*}

\begin{figure*}
\begin{centering}
\includegraphics[width=\textwidth]{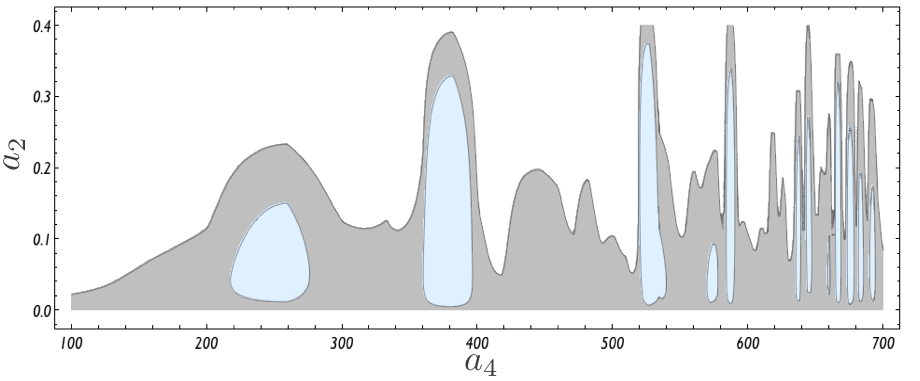}
\par\end{centering}
\caption{(NPH) The $68\%$ and $95\%$ confidence levels for a part of the high frequency grid. Note that the absence of a match between the low and high frequency regimes is artificial. In the high frequency grid, there are only 6 samples for $100\leq a_4 \leq 400$. Additionally, a $100\%$ likelihood was assumed within one grid.  For very high frequencies, the wavelength of the oscillations is too small to be resolved at large angular scale ($l<200$). Primordial oscillations are resolved only in the second peak and beyond. Very high frequencies can produce glitches in the large scale wing of the first peak in the angular power spectrum.  For some specific frequencies and amplitudes these glitches can match outliers in the wing of the first peak of the observed angular power spectrum resulting in $\Delta \chi^2 \simeq 12$ for $a_4 \sim 998$. }
\label{fig:gridcontournph}
\end{figure*}

Given the number of free parameters describing the initial conditions (at least $A_s$, $n_s$ and the frequency and amplitude of the oscillating correction), many computations of the power spectrum are required to obtain the best fit (distribution). A commonly used approach is to apply a Markov Chain Monte Carlo (MCMC), randomly pick values of a set of parameters, and based on the likelihood of the computed run, reject or accept this point to be part of our parameter probability estimate. This approach is highly efficient, as we only compute a limited number of points in the multidimensional likelihood to determine the posterior distribution. Once we add oscillations to the primordial power spectrum an MCMC method becomes less efficient, as the likelihood is expected to become irregular in the coordinate of the oscillation (the frequency) and there are many local maxima in the likelihood function. At some point, we will leave this local maximum and end up in another. Therefore constraints on the initial conditions are hard to recover since the MCMC will constantly move to different local maxima in likelihood space\footnote{Constraints on other parameters unrelated to the initial conditions can be recovered with sufficiently large samples.}.  \cite{Flauger:2009ab} proposed a different approach. Instead of an MCMC over all parameter space, they first considered a grid over a limited number of parameters. The parameters to vary in the grid should be those that determine the primordial conditions (e.g., the frequency). They also identified the baryon density $\Omega_b$ to be degenerate with the amplitude and frequency of the oscillations, as it influences the height of the first peak. Finally they showed that oscillations in axion monodromy models of inflation are a better fit primarily to the first peak in the angular power spectrum, and hence could mimic some of the effects produced by the baryon energy density $\Omega_b$. The grid therefore samples at least 3 parameters (amplitude, frequency and baryon density). The other parameters are fixed to the WMAP 7 \citep{Komatsu:2010fb} best-fit values. As such the null hypothesis (no oscillations) is the fit to beat. The advantage of the grid is that one probes the likelihood completely, although under fixed conditions for most of the $\Lambda$CDM parameters. Under the assumption that these other parameters are not (strongly) correlated with the varying parameters, we should be able to determine the absolute maximum in the likelihood rather than a local one. Once this local maximum has been determined, we can perform an MCMC with the frequency set to the best fit value, and allow all other parameters to vary. This should probe the likelihood of, e.g., the amplitude more efficiently.

We ran three grids, one for the BEFT model and two for the NPH model.  While in this paper we have opted to avoid details about the exact theoretical prediction of the parameters in each model, we decided to consider two grids for the NPH model focusing on two frequency regimes $a_4$. This is motivated by the fact that the NPH model is not expected to have a significant number of oscillations, as the theoretically predicted frequency is slow-roll suppressed, i.e., $a_4\propto \epsilon$. We have therefore investigated low frequencies ($1\leq a_4\leq200$) with 200 log-spaced samples, and a high frequency regime ($100\leq a_4\leq1000$) separated into 500 logarithmically spaced samples. The amplitude $a_2$ runs from $0$ to $0.4$ in 120 and 200 equidistant steps, respectively. For the NPH grid $a_3 = \zeta = 0$. For the low frequency grid we set $0.021\leq \Omega_b h^2 \leq 0.026$ in 10 equidistant steps, resulting in a total of 240.000 grid points, while for the high frequency regime we considered $0.02\leq \Omega_b h^2 \leq 0.027$ in 16 equidistant steps, with a total of 1.6 million grid points. 

\begin{figure*}
\begin{centering}
\includegraphics[scale = 0.44]{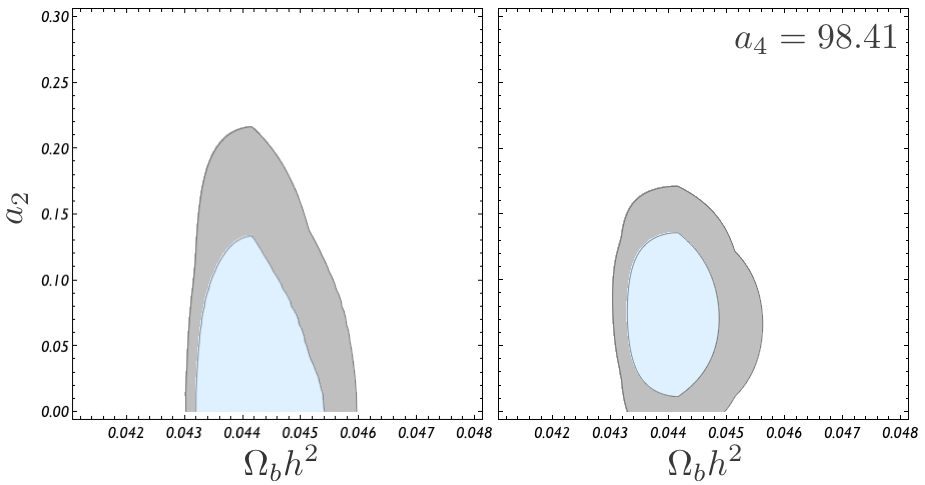}
\par\end{centering}
\caption{(NPH) The $68\%$ and $95\%$ confidence levels of $\Omega_b h^2$ versus $a_2$. The left shows the joint likelihood after marginalizing over the frequency $a_4$ (low frequency grid), while the right shows the joint likelihood for the best fit frequency $a_4 = 98.41$. }
\label{fig:marcontournph}
\end{figure*}

In figures~\ref{fig:gridcontournph2} and \ref{fig:gridcontournph} we show the confidence contours for grids obtained for the NPH model of the frequency versus the amplitude in the low and high frequency regime, marginalized over the baryon energy density $\Omega_b$. Peaks are areas in which the fit is best for non zero values of the amplitude of the modification, while valleys represent frequencies which are not a good fit to the data and the best fit is no modification. For example one can consider a likely frequency (peak) and plot the  probability of the amplitude for that frequency to find that the most likely value for the amplitude is non zero. In fig.~\ref{fig:marcontournph} we show the joint likelihood contour plot for the amplitude $a_2$ and the baryon density $\Omega_b h^2$ marginalized over the frequency $a_4$. We find that $a_2=0$ is the most likely value in the low frequency grid. We also derived the joint likelihood for one of the best-fit frequencies ($a_4 = 98.74$) to show that the best-fit point has a non-zero amplitude $a_2$ with almost $95\%$ CL. 

\begin{figure*}
\begin{centering}
\includegraphics[scale =0.45]{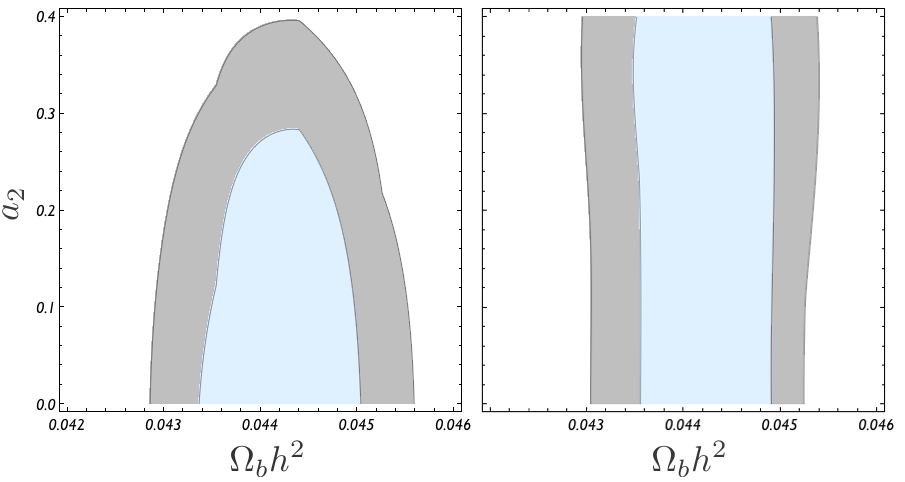}
\par\end{centering}
\caption{(NPH) The $68\%$ and $95\%$ confidence levels of $\Omega_b h^2$ versus $a_2$. The left shows the joint likelihood after marginalizing over the frequency $a_4$ for the high frequency grid up to $a_4 \leq 700$ (63 frequencies). If one includes all frequencies in the marginalization the constraint on $a_2$ disappears (right). }
\label{fig:marcontournph2}
\end{figure*}

For the high frequency regime we show the effect of the highest frequencies on the marginalized amplitude $a_2$ in fig.~\ref{fig:marcontournph2}. Note that the confidence levels are determined assuming that the grid contains all possible values the frequency could have. In other words, one would hope that for either extremely large or small frequencies the likelihood of the fit would go to zero. The problem is that it does not, and therefore we can only determine the confidence levels within a prior determined parameter domain. We have partly motivated this domain on theoretical arguments. Observationally, data is the limiting factor.

It should be obvious that many different frequencies represent good fits. As we had foreseen, this complicates running a large MCMC for {\it all} relevant cosmological parameters. The best fit point  we find for the NPH model ($\Delta \chi^2\sim 12$) centers around very high frequencies, with  $a_4 = 980$ and $a_2 = 0.39$. In fact for this frequency, the best fit amplitude probably lies outside the domain of  $0 \leq a_2 \leq 0.4$. This  implies a relatively large number of primordial oscillations. The angular power spectrum for such high frequencies has most of its oscillations damped, since this frequency is only resolved at scales beyond the first peak. \cite{Martin:2003sg} showed that for these scales the amplitude will be suppressed, similar to how the overall power is damped.  Given the large measurement error at these scales, we would not expect the fit to improve that much. It turns out that at large angular scales, the barely resolved high frequency causes glitches in the large scale wing of the first peak. The WMAP data contains outliers in the large scale (small $l$) wing of the first peak (attempts have been made to understand this in terms of a feature in the primordial spectrum, see e.g. \cite{Covi:2006ci} and recently a principal component analysis has been performed to search for such a feature in \cite{Dvorkin:2011ui}); these are fitted for very specific unresolved frequencies of the primordial power spectrum. These glitches are expected as the angular power spectrum will no longer be able to sample all oscillations appearing in  the primordial power spectrum. We have not been able to get rid of these glitches by increasing the overall accuracy of the numerics of the code. 

In the low frequency regime we obtain two frequencies that give an improvement of $\Delta \chi^2\sim 6$ compared to no oscillations. In the next  section we will discuss the results from an MCMC run for one of these frequencies, with a varying phase and amplitude. For the NPH model, the results can be summarized as follows: for frequencies $1<a_4<200$, $a_2<0.13$ at $68$\% and $a_2<0.21$ at $95$\% confidence. For frequencies up to $a_4 = 700$, $a_2<0.29$ at $68$\% and $a_2<0.39$ at $95$\% while for higher frequencies the amplitude is not constrained within the bounds of the grid. The most likely value within the low frequency grid is $a_4 = 46.5$ with an amplitude of $a_2 \sim 0.056$.

\begin{figure}
\begin{centering}
\includegraphics[scale=.45]{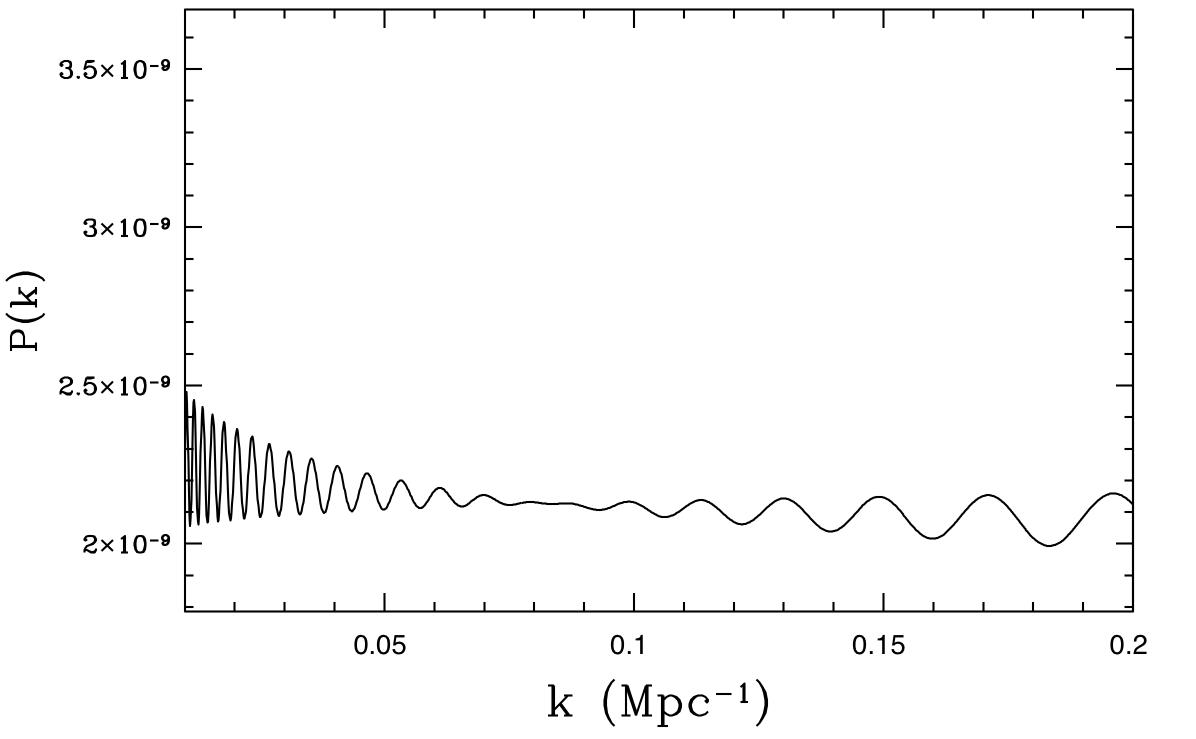}
\par\end{centering}
\caption{The primordial spectrum from the NPH model with $a_4\sim 46$, $a_2 = 0.14$, $a_3 = -0.04$ and $\phi = 1.5$.}
\label{fig:bestfitprim}
\end{figure}

\begin{figure}
\begin{centering}
\includegraphics[scale=.45]{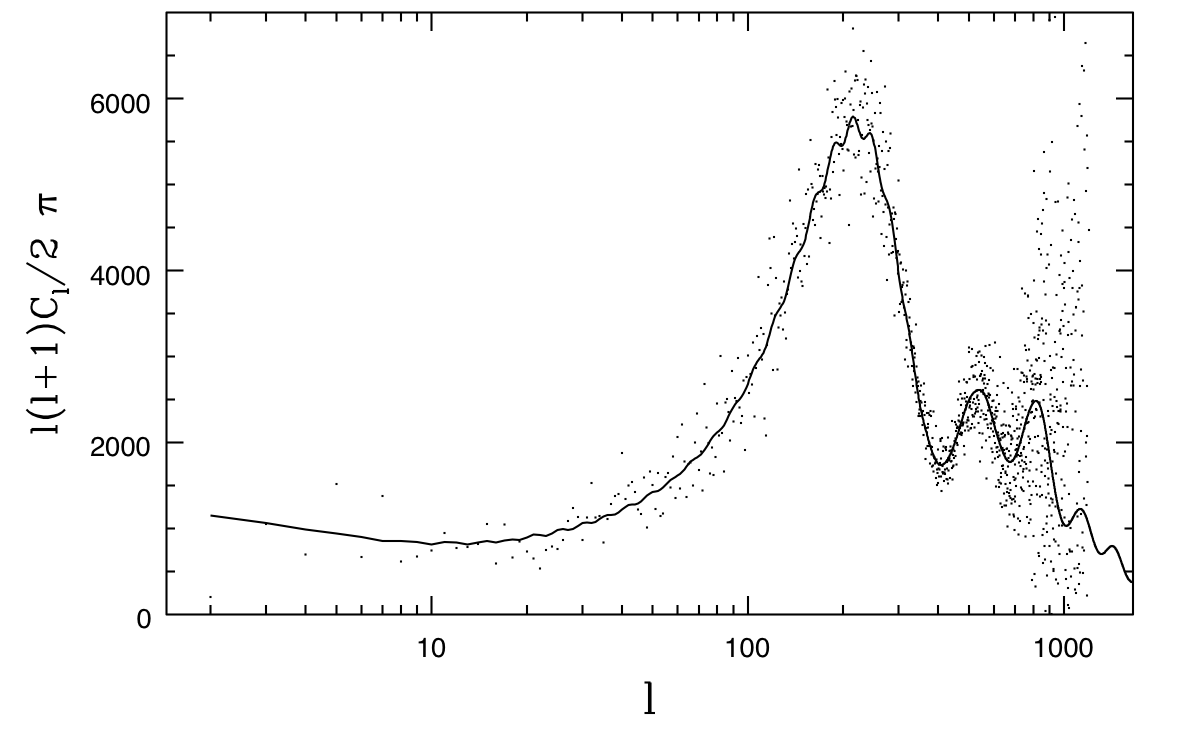}
\par\end{centering}
\caption{Best fit for the NPH model from the MCMC chain, with a fixed frequency $a_4 \sim 46$  . The improvement of the fit compared to no oscillations is $\Delta \chi^2\sim 12$. From the grid we found that an equally good fit is for a frequency $a_4 \sim 98$, about double this frequency. We find that the best fit amplitude is rather large, $a_2 \sim 0.14$, with the $68\%$ level still allowing zero amplitude (fig.~\ref{fig:mar_par_nph}). The amplitude is similar to the best fit amplitude of the axion monodromy model derived in \cite{Flauger:2009ab}. However, this best fit was at a translated frequency of about 150. }
\label{fig:bestfit}
\end{figure}

\begin{figure}
\begin{centering}
\includegraphics[scale=.45]{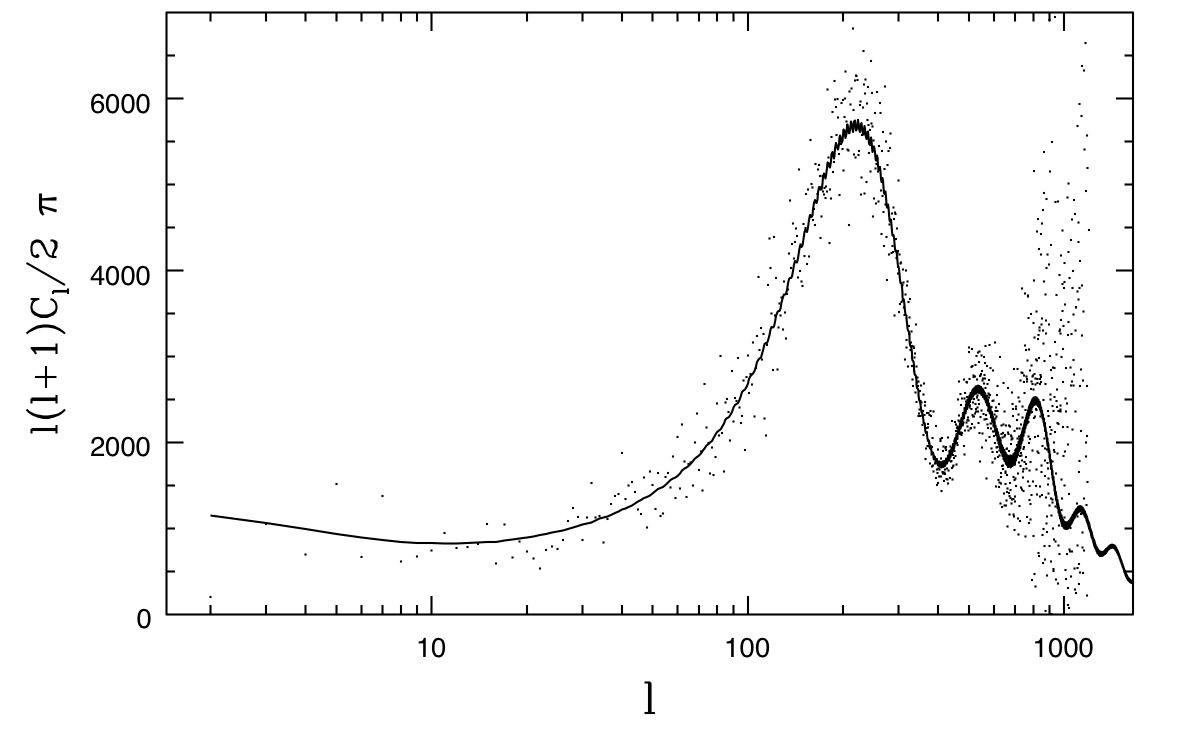}
\par\end{centering}
\caption{The best-fit angular power spectrum derived from the grid for the BEFT model with $A_0 = 10$ and $A_1 = 7708$, resulting in $\Delta \chi ^2 \simeq 13$ relative to no oscillations. }
\label{fig:bestfitprimbeft}
\end{figure}

For the BEFT model, there is a constraint set by BEFT validity on the value of $A_0$. A BEFT approach to initial state modifications is only valid if the physical momentum is smaller than the cut-off scale at the boundary, i.e., $k/a_0<M$. The maximum value of $k$ is set by the smallest observable scale in the CMB, $k_{max} = \mathcal{O}(0.1)$ and therefore we deduce $A_0 \leq10$ from eq.~\eqref{eq:PK_beft}. The number of samples in $A_0$ is set to 120 equidistant values between 1 and 10. For BEFT the frequency is not constrained by slow-roll and is proportional to $M/H$; the scale of new physics divided by the Hubble scale. Consequently the effective frequency can be quite high. We sample 700 logarithmically spaced steps between $10^3$ and $10^4$ making up a total of 1,344,000 grid points. 

\begin{figure*}
\begin{centering}
\includegraphics[width=\textwidth]{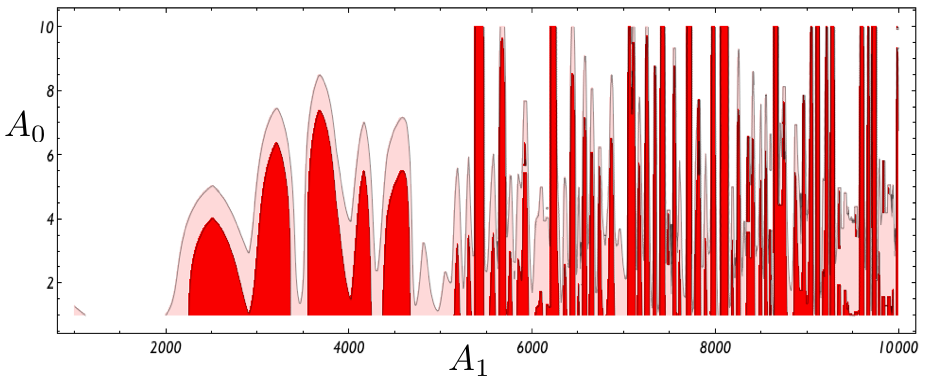}
\par\end{centering}
\caption{(BEFT) The $68\%$ and $95\%$ confidence levels for amplitude $A_0$ versus the frequency $A_1$ marginalized over the baryon density. Again, there are many local maxima in the likelihood. The cut-off in amplitude $A_0$ is set by theoretical constraints. If we would allow larger amplitudes ($A_0>10$), the likelihood contours would likely move towards higher $A$. The  most likely grid point is ($A_1 = 7708$, $A_1 = 10$). }
\label{fig:gridcontourbeft}
\end{figure*}

The likelihood confidence contours for the BEFT model are different from those of the NPH model (fig.~\ref{fig:gridcontourbeft}). Most importantly, the resulting contour does not have a vanishing amplitude, which was assumed as a prior based on the theoretical form of $A_0$. The best fit point in the grid is given by $A_1 = 7708$ with $A_0 = 10$ corresponding to an improvement of $\Delta \chi^2\simeq 13$. 

%
%

\section{MCMC and Model Constraints}\label{sec:MCMC}

\begin{figure*}
\begin{centering}
\includegraphics[width=\textwidth]{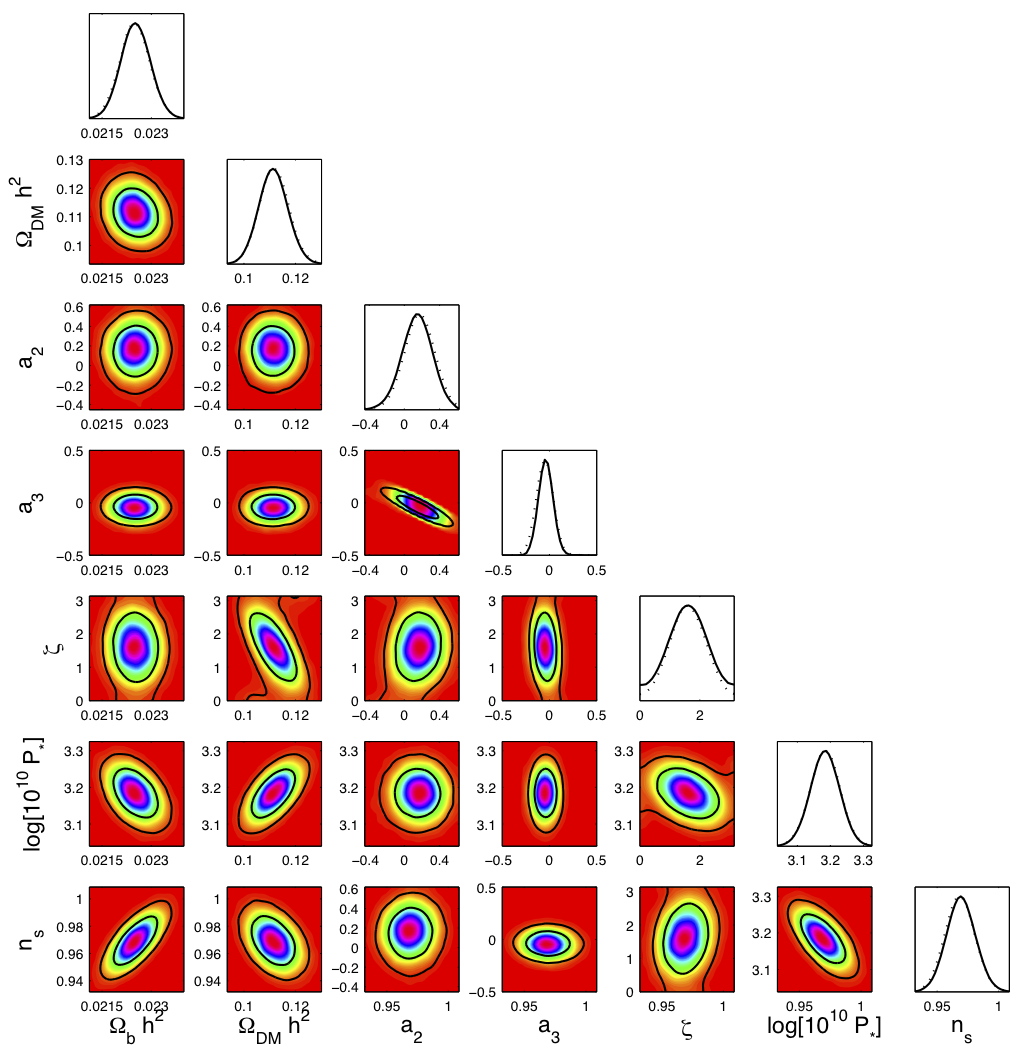}
\par\end{centering}
\caption{Marginalized and joint likelihoods for several parameters in the new NPH scenario. The parameters $a_2$ and $a_3$ show strong correlation.}
\label{fig:mar_par_nph}
\end{figure*}

We analyzed the WMAP data using Monte Carlo Markov sampling with a fixed frequency derived from the grids in the previous section for both BEFT and NPH model. We set a Gelman and Rubin criterion of $R-1 < 0.01$ \citep{GR1992}. The Gelman-Rubin diagnostic $R$ relies on parallel chains to test whether they all converge to the same posterior distribution by considering the variance of the parameters in each chain compared to the variance of the same parameters over all parallel chains.  Convergence is diagnosed once the chains have `forgotten' their initial values, and the output from all chains has become indistinguishable ($R-1=0$). For the NPH model, we ran 8 parallel MCMC with $a_4\sim 45.9$, close to the best fit point in the low frequency grid. In addition to $a_2$ we allowed both $a_3$ and $\zeta$ to vary. We assumed flat priors with $-0.6 \leq a_2 \leq 0.6$, $-0.5\leq a_3 \leq 0.5$ and $0\leq \zeta \leq \pi$. We did not reach our desired GR criterion with the first chains, which is mostly entirely due to the slow convergence of the amplitude $P_*$. After we ran 8 chains, we derived a proposal covariance matrix. We used this covariance matrix to speed-up the convergence. Obviously, if the proposal is derived from chains that have not converged sufficiently, we will not recover the true posterior distribution. As such, using a proposal covariance matrix is not without risks. With the addition of a covariance proposal we easily obtained $R-1<0.003$ over 4 chains. The result for some of the marginalized and joint likelihoods is shown in fig.~\ref{fig:mar_par_nph} for the NPH model. Note the high correlation between $a_2$, the amplitude of the oscillation, and $a_3$, the scale dependence of the oscillatory correction. We found no evidence for a strong correlation between $a_2$ and $a_3$ and the other parameters, which proves that a grid sampling is a very good first estimate of the best fit values of both parameters.  We found that $\zeta$ is weakly correlated with both the amplitude $P_*$ as well as the dark matter energy content of the Universe, explaining the relatively slow convergence of these distributions.  


From the MCMC for the NPH scenario we deduce marginalized best fit values  $a_2 = 0.15\pm 0.17$, $a_4 = 0.04\pm0.04$ and $\zeta = 1.6\pm 0.7$ with a fixed frequency of $a_4 = 45.9$ and a best fit point with $\Delta \chi^2 \simeq 12$. Therefore we conclude that at the 1 sigma level, $a_2$ is consistent with zero.

To relate these to constraints on the NPH model one must investigate the relation between $a_2$, $a_3$ and $a_4$ as derived in \cite{Jackson:2011qg}. The exact form of these parameters is presented in eq.~(31) of their paper\footnote{ We derived these constraints in the assumption $\epsilon_1/\epsilon_2 \sim 1$ and $H_* \sim H$ due to weak scale dependence \citep{Jackson:2011qg}.} We derive a relation between $a_2$, $a_3$ and $a_4$
\begin{eqnarray}
\frac{a_2}{a_3}& \simeq & -  \frac{1}{a_4}\frac{ M}{H} \left[ 5/2+\ln M/H\right]^{-1}.
\end{eqnarray}
Here we took the lower limit $\Lambda = M$, where we consider $M/H>10^2$. The observational limits were derived for $a_4\sim 46$, which puts a bound on $M/H>10^3$ for $\epsilon<0.01$ (slow-roll). This results in a theoretical constraint $a_2/a_3 < - 2.3$. We can derive a similar constraint for the upper limit $\Lambda = \frac{1}{2}(H+M^2/H)$, or $\Lambda/H \sim 1/2(M/H)^2$. It turns out that $a_4$ in this limit can only have a negative sign. For a sine, this means the amplitude picks up a minus sign. We find $M/H>10^3$ and 
\begin{eqnarray}
\frac{a_2}{a_3}& \simeq & \frac{1}{a_4}\frac{ M}{H}(2-\ln M/2H) \left[ 5/2+\ln \frac{1}{2}(M/H)^2\right]^{-1}
\end{eqnarray}
which together results in $a_2/a_3>5.86$. 

\begin{figure}
\begin{centering}
\includegraphics[scale=.38]{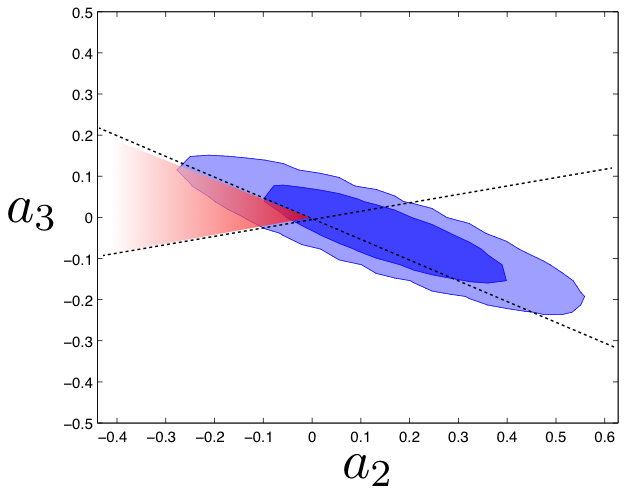}
\par\end{centering}
\caption{Joint likelihood for $a_2$ and $a_3$ together with the theoretical constraints relating $a_2$ to $a_3$ showing $68\%$ and $95\%$ confidence levels. The region (shaded, red) allowed within the NPH model constitutes a small part of the observationally allowed region. It shows that a marginal improvement of the confidence contours could exclude NPH as the source of this oscillation. In addition, both parameters are perfectly consistent with zero.}
\label{fig:joint_constraint}
\end{figure}

In fig.~\ref{fig:joint_constraint} we show the joint likelihood together with the theoretically allowed values of $a_2$  and $a_3$ for a given frequency $a_4$. If this frequency is a valid signal, a minimal improvement of the confidence contours could exclude this oscillation to be due to a NPH altogether.

For the BEFT model, the bound on the frequency is not that stringent and we simply chose the best fit frequency of $A_1 = 7708$ obtained from the grid. Note that this high frequency requires us to compute all $l$ in order to resolve the primordial oscillation. This significantly increases computation time for each power spectrum. From the MCMC we would like to derive proper distributions, which could be an issue if we put priors on $A_0$ that exclude values that have non-zero probability. Therefore we allow $0 \leq A_0 \leq 25$ and $-\pi \leq\phi \leq \pi$. 

The 8 parallel MCMC for the BEFT model resulted in an improvement of $\Delta \chi^2 \sim16$,  with all parameters satisfying the Gelman-Rubin diagnostic, except $A_0$ and $\phi$. We ran 6 additional parallel chains with an estimated covariance matrix from the first 8 chains. The Gelman-Rubin diagnostic was reduced for both $A_0$ and $\phi$ to $R-1< 0.01$.

Again we find no indication that $A_0$ and $\phi$ are strongly correlated with any of the other parameters. In fact, $A_0$ almost seems independent of the other parameters. This is reflected in the fast convergence of the distribution of this parameter, which reached a $R-1< 0.2$ after running the first 8 chains (which is remarkable given the length of the chains). Again, the phase $\phi$ is weakly correlated with all the energy densities, leading to a relatively slow convergence of these distributions, even after estimating the covariance matrix. This correlation however is even less profound here than in the NPH model. These parameters reach $R-1< 0.02$ within the limited running time (6 chains, each 120 hours on 8 core 2.2 Ghz CPU's, resulting in chains of approximately 45000 samples) of the chains. 

For a comparison with theoretical bounds, we derive a relation between $A_0$ and $A_1$ as $A_1\simeq A_0\times (M/H)$ (see \cite{GSSS2005} table~1). For a fixed frequency of $A_1 \sim 7708$ and in the assumption the true distribution is a Gaussian we derive  $7.0\leq A_0 \leq 15.2$ at $68\%$ confidence, which is almost 3 sigma away from zero. From this value we derive that $500 \leq M/H\leq 10^3$ at $68\%$ confidence which lies within theoretical predictions. Based on the best fit amplitude $A_0\sim13$ and the average likelihood we obtain a slightly different constraint of consequently we obtain $450 \leq M/H\leq 850$ at $68\%$ confidence. We argued that in principle $A_0$ is constrained to be smaller than 10 in the BEFT model.  We set the coupling constant $\beta = 1$ but in fact $\beta$ is $\mathcal{O}(1)$ which allows for small increase of the total amplitude, $\beta A_0$, possibly matching the best fit amplitude. Unlike the best-fit oscillation from the NPH model, this correction is inconsistent with zero at the 3 sigma level.

\section{Effects on Other Parameters}
\begin{figure*}
\begin{centering}
\includegraphics[width=\textwidth]{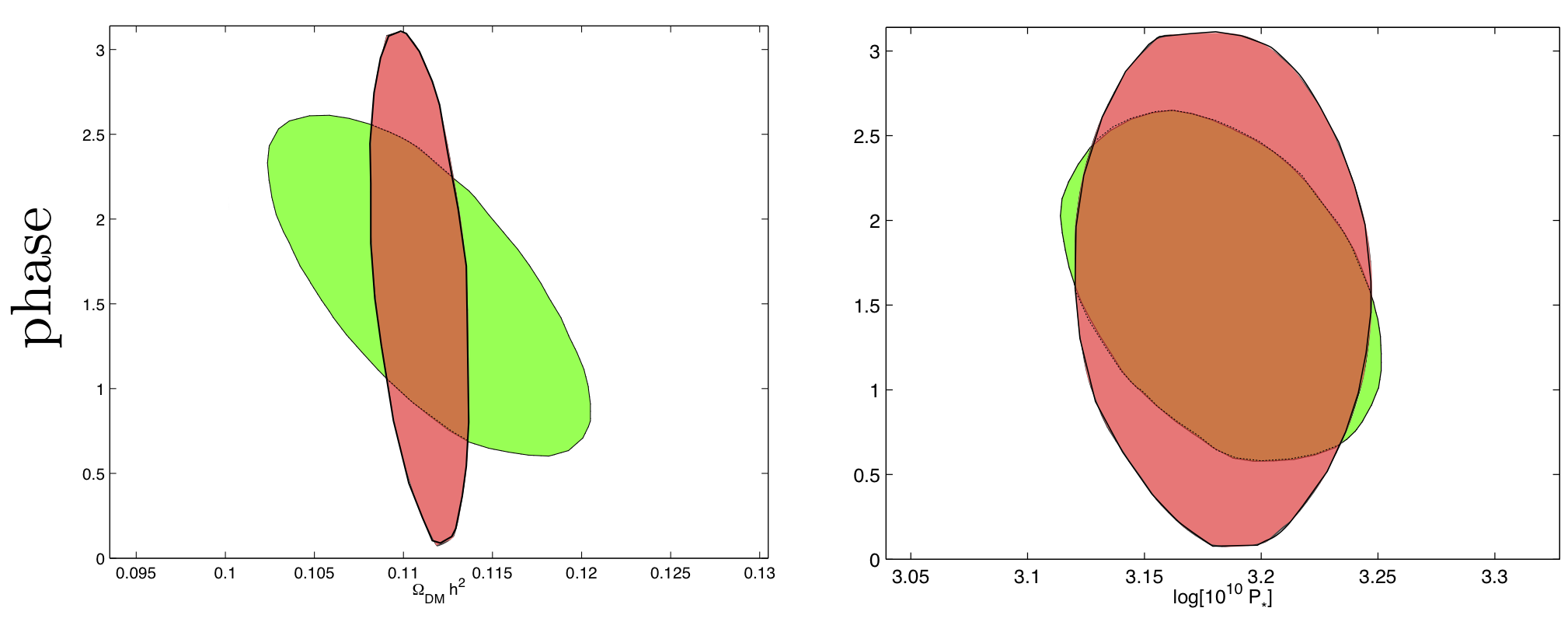}
\par\end{centering}
\caption{The $68\%$ confidence contours between the phase $\zeta$ (NPH, green) and $\phi$ (BEFT, red) and the dark matter energy density and the amplitude. For sake of comparison we shifted the central value of $\phi$ upwards, while contour levels were maintained. Both examples show how the correlation between $\zeta$ and $P_*$ and $\Omega_{DM}$ affects the uncertainty in these parameters. It does seem that overall uncertainty is conserved, as the error in $\phi$ is larger than the error in $\zeta$.}
\label{fig:contours_omb_amp}
\end{figure*}

Although there is little to no correlation between most of the parameters from the oscillating component and the other parameters within $\Lambda$CDM, the cross correlation between $\Lambda CDM$ parameters is affected differently for BEFT and NPH (the best fit values however are similar). For example, although confidence the levels of $n_s$ are hardly affected by the presence of oscillations, in the NPH model, weak correlation between the phase and the energy densities causes the probability distribution in $\Omega_{DM}$ to broaden, i.e., the correlation between the phase and the energy densities in the NPH scenario causes the uncertainty in those parameters to increase. This effect is shown in fig.~\ref{fig:contours_omb_amp}.

\section{Conclusions}\label{sec:conclusions}

We used the latest WMAP data to constrain oscillations on top of an almost scale invariant primordial power spectrum. We argued that the primary difficulty in constraining these models with the data is related to the irregular likelihood function of the frequency of these oscillations. There are many equally good fits and these fits are at discrete frequencies. For all other parameters one tries to constrain in a $\Lambda$CDM model of the Universe, such degeneracy usually does not exist and it is quite sufficient to scan parameter space using an MCMC. In order to avoid the random jumps from one maximum in the likelihood to another, it is preferable to keep the frequency fixed. However, the frequency is on of the key parameters being constrained, and it would make no sense to fix it prior to analyzing the data. In order to estimate the frequency before the MCMC we applied a grid search for the best fit. This grid quickly becomes incalculable for a large number of parameters (which is the primary reason to run an MCMC instead) and we only varied the parameters characterizing the modification of the power spectrum as well as the baryon density, which previously had been identified to be correlated with the parameters of the oscillatory correction. Once the best fit had been established we performed an MCMC with the frequency fixed to its best fit value. 

We found that the addition of oscillations on top of the smooth power spectrum can improve the overall fit. An improvement up to $\Delta \chi^2 \sim16$ was found once we applied an MCMC with a predetermined best fit value of the frequency, which resulted in a best fit amplitude almost 3 sigma away from zero. We did not find significant correlation between the oscillatory parameters {\it besides} the frequency and the parameters in $\Lambda$CDM, confirming that grid sampling is a fairly robust first estimate of the best-fit frequency. 

In the NPH scenario very high frequencies should lead to unobservable amplitudes. For completeness we analyzed such high frequencies. We found that for some specific frequencies we could obtain an improvement in the goodness of fit up to $\Delta \chi^2 \sim 12$ compared to no oscillations. Analyzing the angular power spectrum with these extreme primordial frequencies revealed that these frequencies are in fact unresolved at large angular scales (small $l$), where the wavelength of the oscillating primordial power spectrum is smaller than the angular distance between subsequent $l$'s. This leads to small glitches in the slope of the first peak of the angular power spectrum, which fit some of the observed outliers in the slope of the first peak in the observed power spectrum. As such, high frequencies, although unresolved, could account for some of the large scale effects we observe in the data, and  improve the overall fit. We conclude, however, that these oscillations cannot be caused by NPH modifications and must be due to a different model, possibly an axion-monodromy inflation type model \citep{Flauger:2009ab}.

Although in the grid for the low frequency domain $1\leq a_4 \leq 200$ of the NPH model we could only found one frequency with an improvement of $\Delta \chi^2 \simeq 6$, we ran an MCMC near this frequency to obtain an improvement of $\Delta \chi^2 \simeq 12$ corresponding to a best-fit amplitude of $a_2=0.16$. This is close to the best-fit amplitude found in \cite{Flauger:2009ab} with an improvement of $\Delta \chi^2 \simeq 11$ (WMAP-5). Their improvement appeared at a (translated) frequency of $a_4 \sim 150$. It is interesting to note that there appear to be subsequent improvements at low frequencies close to $50$, $100$ (WMAP-7) and $150$ (WMAP-5) which could be an additional hint we might be looking at an oscillation as opposed to noise, because the improvement appears at equidistant intervals in frequency.

Using an MCMC we have been able to put constraints on several primordial parameters. In particular, we derived constraints on the amplitude $a_2$, the tilt $a_3$ and the phase $\zeta$ of the oscillatory correction in the NPH model. We used these observational constraints to test the NPH model by implementing the relation between various parameters as predicted by the NPH model. This shows that the signal found at this frequency could originate from NPH modifications to the primordial power spectrum. However, an improvement of the confidence levels would exclude this possibility.  

For the BEFT model we ran one grid and we found that even before varying all parameters in an MCMC we could achieve an improvement of $\Delta \chi^2 \simeq13$. This improvement corresponded to a frequency $A_1 = 7708$ and an amplitude $A_0 =10$. 
We ran an MCMC around this frequency and found that the theoretical bound on $A_0$ does not probe the observationally best fit point. In order to recover the full distribution of $A_0$ we set an upper limit on $A_0<25$ when running the MCMC. We found a best fit improvement of  $\Delta \chi^2 \sim16$ compared to no oscillations with a phase $\phi = -0.21$ and an amplitude of $a_2 = 12.25$. This best fit value of  $A_0$ is theoretically not expected by a BEFT modification of the primordial power spectrum. The amplitude of the oscillatory correction is also proportional to a coupling constant $\beta$ which we set to $1$. Once we relax this assumption, larger amplitudes are theoretically possible through $\beta A_0$. These results enabled us to put a constraint on the ratio $M/H$ of $5\times 10^2\lesssim M/H \lesssim 10^3$, which is a very realistic possibility of this parameter ratio.   

It is hard to assess the significance of the improved fits. We could think of two possibilities: investigate simulated CMB data without oscillations with realistic beam and noise, and investigate simulated CMB data {\it with} oscillations and realistic beam and noise. For example, if we can find oscillations in simulated data without oscillations, this would suggest oscillations can well be mimicked by noise. Complementarily, generating realistic CMB data including primordial oscillations, can we recover these oscillations by analyzing the data? In particular, it would be interesting to determine whether there exists a threshold amplitude for oscillations to be recovered. Preferably we should generate a large number of maps in order to see if we can assess a probability of recovering oscillations from the data. We will report our findings in future work. 

For now, we conclude that a primordial power spectrum with no oscillations is consistent with the data for most amplitude and frequency of the primordial signal.  For some frequencies, a non-zero amplitude of the oscillatory corrections to the power spectrum seems to be preferred by the data. These signals can be investigated, and used to constrain primordial parameter space, possibly signifying some of the detailed physics driving inflation. 

\medskip

{\large{\bf{Acknowledgments}}}
We would like to thank Mark Jackson for very helpful discussions and for reading the proof manuscript. The authors would also like to thank Christophe Ringeval who provided some of the essential code which was used throughout this paper. PDM would like to thank Yuri Cavecchi and  Marcin Zielinski for helping setting up the code on the lisa super cluster. PDM was supported by the Netherlands Organization for Scientific Research (NWO), NWO-toptalent grant 021.001.040. The research of JPvdS is financially supported by Foundation of Fundamental Research of Matter (FOM) research grant 06PR2510.

\bibliographystyle{apj}

 \end{document}